\documentclass{eas}
\usepackage{graphicx}
\def\lesssim{{_ <\atop{^\sim}}}

\def\ap3m{AP$^3$M}
\def\LCDM{$\Lambda$CDM}

\def\hMpc{$h^{-1}{\ }{\rm Mpc}$}
\def\hMsun{$h^{-1}{\ }{\rm M_{\odot}}$}

\def\nbody{$N$-body}
\def\c15{$c_{\rm 1/5}$}

\def\vecx{\vec{x}}
\def\vecp{\vec{p}}
\def\vecr{\vec{r}}
\def\vecF{\vec{F}}
\def\vecg{\vec{g}}

\def\vecnabla{\vec{\nabla}}

\newcommand{\Eq}[1]{Eq.~(\ref{#1})}
\newcommand{\Fig}[1]{Fig.~\ref{#1}}
\newcommand{\amiga}{\texttt{AMIGA}}
\def\ea{et~al.~}                            
\def\lesssim{\mathrel{\hbox{\rlap{\hbox{\lower4pt\hbox{$\sim$}}}\hbox{$<$}}}}
\def\gtrsim{\mathrel{\hbox{\rlap{\hbox{\lower4pt\hbox{$\sim$}}}\hbox{$>$}}}}

\newcommand{\ApJ}[3]    {\mbox{#3, ApJ, #1, #2}}
\newcommand{\ApJS}[3]   {\mbox{#3, ApJ, #1, #2}}

\newcommand{\ARAA}[3]   {\mbox{#3, ARAA, #1, #2}}

\begin{document}

\TitreGlobal{Mass Profiles and Shapes of Cosmological Structures}

\title{MONDian Cosmological Simulations}
\author{Knebe, Alexander}\address{Astrophysikalisches Institut Potsdam, An der Sternwarte 16, 14482 Potsdam, Germany}
\runningtitle{MONDian cosmological simulations}
\setcounter{page}{1}
\index{Knebe, Alexander}

%
\begin{abstract} 
We present results derived from a high-resolution cosmological \nbody\ simulation in which the equations of motion have been changed to account for MOdified Newtonian Dynamics (MOND). It is shown that a low-$\Omega_0$ MONDian model with an appropriate choice for the normalisation $\sigma_8$ of the primordial density fluctuations can lead to similar clustering properties at redshift $z=0$ as the commonly accepted (standard) \LCDM\ model. However, such a model shows no significant structures at high redshift with only very few objects present beyond $z>3$. For the current implementation of MOND density profiles of gravitationally bound objects at $z=0$ can though still be fitted by the universal NFW profile.
\end{abstract}

\maketitle

%
\section{Introduction}
Although the currently favoured \LCDM\ model has proven to be remarkably successful on large scales (cf. Spergel et al. 2003), recent high-resolution \nbody\ simulations seem to be in contradiction with observation on sub-galactic scales: the CDM "crisis" is far from being over. Suggested solutions to this include the introduction of self-interactions into collisionless \nbody\ simulations (e.g. Spergel~\& Steinhardt 2000), replacing cold dark matter with warm dark matter (e.g. Knebe~\ea 2002) or non-standard modifications to an otherwise unperturbed CDM power spectrum (e.g. bumpy power spectra, Little, Knebe~\& Islam 2003). Some of the problems, as for instance the overabundance of satellites, can be resolved with such modifications but none of the proposed solutions have been able to rectify \textit{all} shortcomings of \LCDM\ simultaneously. Therefore, alternative solutions are unquestionably worthy of exploration, one of which is to abandon dark matter completely and to adopt the equations of MOdified Newtonian Dynamics (MOND; Milgrom 1983; Bekenstein~\& Milgrom 1984).

\begin{table}
\caption{Model parameters. In all cases a value for the Hubble parameter
        of $h=0.7$ was employed.}
\label{parameter}
\begin{center}
\begin{tabular}{lllllll}

label   & $\Omega_0$ & $\Omega_b$ & $\lambda_0$ & $\sigma_8^{z=0}$ & $\sigma_8^{\rm norm}$ & $g_0$ [cm/s$^2$]\\ 
\hline \hline
\LCDM\  &    0.30    &   0.04    & 0.7  &   0.88    & 0.88  & ---\\
OCBM    &    0.04    &   0.04    & 0.0  &   0.88    & 0.88  & ---\\
OCBMond &    0.04    &   0.04    & 0.0  &   0.92    & 0.40  & 1.2 $\times 10^{-8}$ \\

\end{tabular}
\end{center}
\end{table}

\section{The Simulations}
We adapted our cosmological \nbody\ code \amiga\footnote{\texttt{http://www.aip.de/People/AKnebe/AMIGA}} (Knebe, Green~\& Binney 2001) to account for the effects of MOND in the following way. In an \nbody\ code one usually integrates the (comoving) equations of motion

\begin{equation}\label{eomPec}
  \dot{\vecx} =   {\vecp \over a^2} \ , \ \ \dot{\vecp} =  {\vecF_{\rm pec}\over a} 
\end{equation}

\noindent
which are completed by Poisson's equation 

\begin{equation}\label{PoissonPec}
  \displaystyle \vecnabla_x \cdot \vecF_{\rm pec}(\vecx) 
    = -\Delta_x \Phi(\vecx) = - 4 \pi G (\rho(\vecx) - \overline{\rho}) \ .
\end{equation}

In these equations $\vecx = \vecr/a$ is the comoving coordinate, $\vecp$ the canonical momentum, $\vecnabla_x \cdot$ the divergence operator ($\Delta_x$ the Nabla operator) with respect to $\vecx$ and $\vecF_{\rm pec}(\vecx) = -\nabla \Phi(\vecx)$ the \textit{peculiar acceleration field in comoving coordinates}. We now need to modify these (comoving) equations to account for MOND.

Sanders~\& McGaugh (2002) showed that the relation between the Newtonian $g^{\rm prop}$ and MONDian $g_M^{\rm prop}$ acceleration (in proper coordinates) can be written as

\begin{equation} \label{ggN}
 g_M^{\rm prop} = g^{\rm prop} \left(\frac{1}{2} + \frac{1}{2} \sqrt{1+\left(\frac{2 g_0}{g^{\rm prop}}\right)^2}\right)^{1/2} \ .
\end{equation}

\noindent
where we already faciliated Milgrom's interpolation function $\mu(x) = x (1+x^2)^{-1/2}$ (Milgrom 1983) and $g_0$ is the fundamental acceleration of the MOND theory.

If we further assume that MOND only affects peculiar acceleration (in proper coordinates), i.e. $\vecg_{\rm pec}^{\rm prop} = {\vecF_{\rm pec} / a^2}$, the recipe for adding the MOND formalism to an \nbody\ code reads as follows: (1) solve \Eq{PoissonPec} using \amiga\ which gives the comoving $\vecF_{\rm pec}$, (2) calculate the peculiar acceleration in proper coordinates $\vecg_{\rm pec}^{\rm prop} = \vecF_{\rm pec}/a^2$, (3) use \Eq{ggN} to calculate $\vecg_{M, \rm pec}^{\rm prop}$ from $\vecg_{\rm pec}^{\rm prop}$, (4) transfer $\vecg_{M, \rm pec}^{\rm prop}$ back to $\vecF_{M, \rm pec} = a^2 \vecg_{M, \rm pec}^{\rm prop}$, and (5) use $\vecF_{M, \rm pec}$ rather than $\vecF_{\rm pec}$ for the equations of motion~(\ref{eomPec}). For a more elaborate discussion of the assumptions upon which this scheme is based and a more detailed derivation of the formulae presented here we refer the reader to Knebe~\& Gibson (2004).

Our suite of simulations now consists of a standard \LCDM\ model, an open, low-$\Omega_0$ model with the same $\sigma_8$ as \LCDM\ (OCBM), and an open, low-$\Omega_0$ model with MOND and adjusted $\sigma_8$ (OCBMond), and their physical parameters are summarized in Table~\ref{parameter}. We simulated $128^3$ particles in a box of side length 32\hMpc\ from redshift $z=50$ to $z=0$. Gravitationally bound objects were identified using the \amiga's native halo finder \texttt{AHF}  described in great detail in Gill, Knebe~\& Gibson (2004).

\begin{figure}[h]
   \centering
   \includegraphics[width=12.5cm]{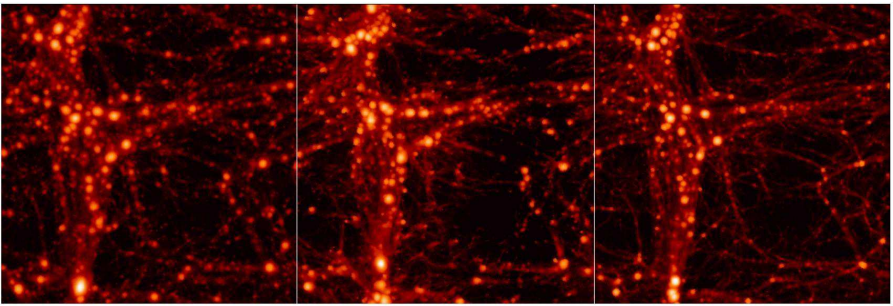}
      \caption{Density field of the
               \LCDM\ (left), OCBMond (middle), and OCBM (right) model
               at redshift $z=0$.}
       \label{figure1}
   \end{figure}

\section{The Results}
The first intriguing result can be viewed in \Fig{figure1} where we show a projection of the whole simulation with each individual particle color-coded according to the local density at redshift $z=0$. This figure demonstrates that the MOND model  exhibits fairly similar features in terms of the locations of high density peaks, filaments and the large-scale structure when being compared to both the standard \LCDM\ model and the OCBM simulation. One should bear in mind though that the OCBMond simulation was started with a much lower $\sigma_8^{\rm norm}$ normalisation than the other two run indicating a faster growth of structures in MOND universes.

\begin{figure}[h]
   \centering
   \includegraphics[width=10cm]{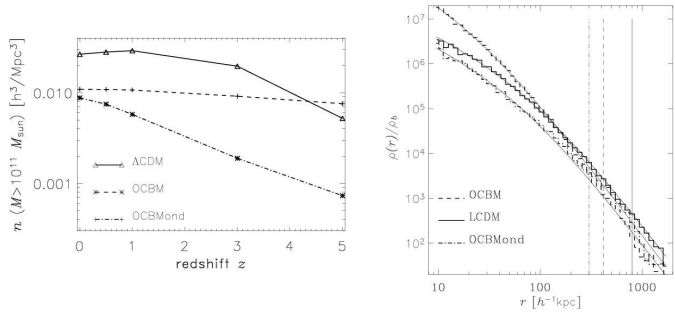}
      \caption{Redshift evolution of the abundance of objects 
               with mass $M>10^{11}$\hMsun\ (left) and the 
               density profile for the most massive halo in all three models (right).}
       \label{figure2}
   \end{figure}

This result is supported by \Fig{figure2} (left panel) where the abundance evolution of gravitationally bound objects more massive than $M>10^{11}$\hMsun\ is shown. Moreover, this figure also poses a serious problem for cosmological MONDian structure formation. Due to the increased formation rate of objects we had to lower the amplitude of the primordial density perturbations. This in turn leads to a strong deficiency of gravitionally bound structures at redshifts $z \geq 3$, quite in contrast to observations where we find galaxies out to redshifts of $z\sim 6$ (Shioya~\ea 2005).

\Fig{figure2} (right  panel) also shows the spherically averaged density profile of the most massive halo in all three models along with fits (thin solid lines) to a Navarro, Frenk~\& White (1997) profile $\rho(r) \propto ((r/r_s) (1+r/r_s)^2)^{-1}$. We observe that even for the OCBMond model the data is equally well described by the functional form of a NFW profile (at least out to the virial radius indicated by the respective vertical lines). However, the central density of that halo in the OCBMond model is lower than in \LCDM\ and especially in OCBM.  A quantitative analysis further shows that the concentration of the OCBMond halo is about a factor of four smaller than in the \LCDM\ model.

\section{The Conclusions}
Even though it is possible to match a cosmological simulation including the effects of MOND to the standard \LCDM\ structure formation scenario at redshift $z=0$ there are serious deviations at higher redshifts. We conclude that the most distinctive feature of a MONDian universe is the late epoch of galaxy formation. However, the density profiles of gravitationally bound objects still follow the universal NFW shape.



\end{document}